\documentclass[12pt]{article}
\usepackage[T1]{fontenc}
\usepackage[utf8]{inputenc}
\usepackage[margin=1in]{geometry}
\usepackage{graphicx}
\usepackage{tabularx}
\usepackage{booktabs}
\usepackage{array}
\usepackage{ragged2e}
\usepackage{wrapfig}
\usepackage{caption}
\usepackage{amsmath}
\usepackage{hyperref}
\hypersetup{
  colorlinks=true,
  linkcolor=blue,
  citecolor=blue,
  urlcolor=blue
}
\usepackage{authblk}
\newcolumntype{P}[1]{>{\RaggedRight\arraybackslash}p{#1}}
\newcolumntype{C}[1]{>{\Centering\arraybackslash\hspace{0pt}}p{#1}}
\title{OpenEye: A Scalable Open-Source Hardware Accelerator for DNNs}
\author[1]{Denis Lebold}
\author[1,2]{Hendrik Wöhrle}
\affil[1]{University of Duisburg-Essen, Campus Duisburg, Bismarckstraße 81, 47057 Duisburg, Germany}
\affil[2]{Fraunhofer Institute for Microelectronic Circuits and Systems, Duisburg, Germany}
\affil[ ]{\texttt{\{denis.lebold, hendrik.woehrle\}@uni-due.de}}
\date{}
\begin{document}
\maketitle
\begin{abstract}
The increasing computational complexity of deep neural network inference poses
significant challenges for efficient hardware acceleration on embedded
platforms, particularly with respect to resource consumption and scalability.
This work presents OpenEye, a scalable and sparsity-aware FPGA-based hardware
accelerator designed to efficiently execute common neural network operations
such as convolutions, dense layers, and pooling.

OpenEye is based on a highly parameterizable architecture composed of clusters
of processing elements interconnected by a streaming-based dataflow. The
paper provides a detailed explanation of the internal operation of the
accelerator, including data movement, buffering strategies, control logic, and
the coordination between clusters and PEs. The architecture natively supports
sparse weights and activations, enabling the efficient processing of sparse data
without unnecessary computations or memory accesses.

A key design property of OpenEye is its scalability: the number of clusters and
processing elements can be varied to adapt the accelerator to different
performance and resource constraints. The design achieves a near-linear scaling
of routing and interconnect overhead with increasing PE counts, which is
essential for maintaining efficiency on large FPGA devices.

To evaluate scalability across different design points, multiple OpenEye
configurations with varying cluster and PE sizes were implemented on a Xilinx
ZU19EG FPGA. Representative neural network operations, including convolutional,
fully connected, and pooling layers, were used to analyze resource utilization,
execution latency, and scalability behavior. The results show favorable
trade-offs between performance and resource consumption across the explored
configurations.

\end{abstract}
\section{Introduction}

The rapid adoption of convolutional neural networks (CNNs) across computer vision, robotics, and embedded intelligence has driven extensive research into dedicated hardware accelerators that outperform general-purpose processors in throughput and energy efficiency. A key insight from prior work is that data movement, not arithmetic computation, increasingly dominates both execution time and energy consumption in neural network inference. This trend arises from large model sizes, deep layer structures, and limited on-chip memory, which together force frequent off-chip memory accesses that are significantly more expensive than on-chip computation~\cite{horowitz2014energy}. Traditional accelerator evaluations often isolate the compute core performance while abstracting away communication overheads or assuming idealized transmission models, thereby underestimating real system costs.

Early spatial architectures such as Eyeriss highlighted the importance of minimizing high-cost data transfers by exploiting data reuse through optimized dataflows like the Row-Stationary (RS) mapping~\cite{eyeriss2017,eyerissv2}. Eyeriss v2 further extended this concept by supporting sparse data execution in the compressed domain and a flexible hierarchical network-on-chip (NoC) that adapts to layer shapes and data reuse patterns~\cite{eie2016,cnvlutin2016}. These innovations demonstrated that addressing dataflow and network interconnect flexibility is essential for high throughput and energy efficiency across diverse CNN models. Nevertheless, most such architectures are evaluated at the ASIC level, often ignoring the detailed cost of interfacing and memory access on reconfigurable platforms such as FPGAs, where interface overhead and memory access behavior can materially impact performance and energy behavior~\cite{finnr2018}.

In parallel, research into sparse accelerators and structured sparsity has shown that skipping zero-valued operations and compressing activations/weights contribute significantly to energy reduction and performance improvement when hardware is designed to exploit these patterns efficiently~\cite{brainwave}. However, many sparse accelerator designs focus on theoretical acceleration potential without fully characterizing communication latency, external memory traffic, and their energy impact. This gap is particularly salient in FPGA implementations, where reconfigurability and memory interface design trade-offs can overshadow computational gains unless carefully accounted for.

To address these limitations, our work presents a hardware accelerator designed to process complete CNN layers within a single transmission cycle, minimizing external memory accesses while supporting scalable sparsity and configurable network shapes. Inspired by flexible and sparse accelerators such as Eyeriss v2, our design targets a more comprehensive evaluation that includes transmission time, memory interface overheads, and energy costs — metrics critical for real-world deployment on both ASIC and FPGA platforms. By bridging the gap between idealized compute-centric evaluations and holistic system-level performance metrics, this work aims to deliver a more realistic assessment of hardware acceleration potential for modern NNs.

\subsection{FPGAs as Target Platforms}

While many state-of-the-art neural network accelerators are evaluated as ASIC designs, FPGA-based implementations play a crucial role in bridging architectural research and real-world deployment. FPGAs enable rapid prototyping, architectural exploration, and system-level evaluation without the high cost and long development cycles associated with ASIC fabrication. More importantly, FPGA platforms expose realistic memory hierarchies and interface constraints, making them particularly well suited for evaluating data movement, transmission overhead, and energy consumption beyond the compute core.

In contrast to idealized ASIC evaluations, FPGA-based systems require explicit consideration of memory interfaces, buffering strategies, and bandwidth limitations. These factors are especially relevant for sparse accelerators, where irregular data access patterns can significantly influence performance and energy efficiency. By implementing OpenEye on FPGA, this work captures the interaction between computation, dataflow, and communication more accurately, providing insights that are difficult to obtain from purely analytical or ASIC-only studies.

Furthermore, the reconfigurability of FPGAs enables to realize different architectural configurations, such as processing element array dimensions, sparsity handling strategies, and memory configurations. This makes FPGA platforms an effective vehicle for evaluating scalable accelerator architectures like OpenEye and for identifying design trade-offs that inform both FPGA- and ASIC-oriented implementations.

\subsection{Related Work}

Eyeriss~\cite{eyeriss2017} introduced the Row-Stationary (RS) dataflow to minimize data movement energy by maximizing reuse through local scratch pads and a hierarchical NoC, demonstrating energy-efficient CNN processing on a 168-PE ASIC. While Eyeriss pioneered dataflow optimization for energy efficiency, OpenEye extends this concept to FPGAs with fully scalable cluster and PE configurations that adapt to different FPGA sizes and layer requirements, while explicitly accounting for external interface and memory bandwidth costs that Eyeriss abstracts away in ASIC simulation.
Eyeriss v2~\cite{eyerissv2} advanced the original design by supporting DNNs through a flexible hierarchical mesh NoC and native sparse processing using compressed sparse column format for both weights and activations. OpenEye builds upon these sparse processing concepts but provides independent X- and Y-dimension scalability in the PE cluster and implements a more flexible routing infrastructure with configurable diagonal activation streaming, enabling arbitrary stride configurations without hardware modification, which Eyeriss v2's fixed mesh topology cannot achieve.
EIE~\cite{eie2016} focused exclusively on accelerating compressed sparse NNs by performing inference directly on compressed weight and activation data, achieving significant energy savings through weight sharing and activation sparsity exploitation. Unlike EIE's fixed architecture optimized only for fully-connected layers, OpenEye supports both convolutional and fully-connected layers with dynamic sparsity handling while maintaining scalability across different network topologies and providing configurable parallelism that EIE lacks.
Cnvlutin~\cite{cnvlutin2016} proposed ineffectual-neuron-free computing to dynamically skip zero-valued activations in convolutional layers, improving energy efficiency by gating off unnecessary computations. While Cnvlutin only exploits activation sparsity, OpenEye natively supports sparsity in both activations and weights through dedicated sparse encoding in address and data RAMs within each PE, providing more comprehensive sparsity exploitation with lower overhead.

FINN-R~\cite{finnr2018} provides an end-to-end framework for deploying quantized NNs on FPGAs, emphasizing binary and low-precision networks to maximize throughput and minimize resource usage. In contrast to FINN-R's focus on extreme quantization and streaming architectures optimized for small networks, OpenEye targets full-precision and moderate-precision networks with scalable cluster-based architectures that support larger models through configurable on-chip buffering and flexible dataflow, making it suitable for a broader range of CNN applications.
Pflow~\cite{pflow2024} presents a heterogeneous acceleration framework for CNNs on FPGAs with dataflow-based operator scheduling and graph optimization for layer-wise deployment. While Pflow emphasizes compile-time graph optimization and heterogeneous computing integration, OpenEye focuses on scalable hardware architecture with runtime-configurable dataflow through its hierarchical router system, providing finer-grained control over data movement and enabling more efficient exploitation of data reuse patterns across diverse layer configurations.
VCONV~\cite{vconv2024} implements a CNN accelerator for FPGAs using conventional systolic array architectures with emphasis on memory bandwidth optimization and on-chip buffering strategies. OpenEye differentiates itself through its hierarchical cluster-based organization with independent cluster and PE scaling, sparse data support, and structured multi-router architecture that enables more flexible data movement compared to VCONV's fixed systolic dataflow.
The Ultra-Low Power FPGA CNN accelerator~\cite{ultralowpowerfpga2024} targets power-constrained edge deployment with optimized data scheduling and resource allocation for specific network architectures. Unlike this work's fixed architecture optimized for specific networks and power constraints, OpenEye provides parametric scalability that enables deployment across different FPGA platforms and power budgets while maintaining consistent architectural principles and sparse data processing capabilities.
The Sparse FPGA CNN accelerator~\cite{sparsefpga2025} implements pattern-compressed sparse neural network processing on FPGAs using structured sparsity patterns to reduce memory access and computation. While this design exploits structured sparsity patterns, OpenEye supports both structured and unstructured sparsity through its flexible sparse encoding mechanism and provides superior scalability through independent cluster and PE configuration, enabling adaptation to networks with varying sparsity characteristics without hardware redesign.
The Fast Readout FPGA CNN accelerator~\cite{fastreadoutfpga2023} optimizes data readout bandwidth and implements multiplier sharing to improve resource utilization in FPGA-based DCNN processing. OpenEye achieves similar bandwidth optimization through its dedicated router architecture and Ultra-RAM buffering strategy while additionally providing scalable sparse processing and flexible cluster interconnection that enables processing of complete feature maps on-chip, reducing external memory traffic beyond what multiplier sharing alone can achieve.

Table~\ref{tab:related_work} summarizes the comparison of OpenEye with these state-of-the-art accelerators. OpenEye uniquely combines scalable sparse execution, flexible hierarchical dataflow with configurable routers, and comprehensive evaluation of memory and interface costs on real FPGA hardware, addressing limitations present in both ASIC-focused designs that abstract away interface costs and FPGA designs that lack scalable sparse dataflow architectures.
\setlength{\tabcolsep}{0pt}
\begin{table*}[t]
\begin{tabularx}{\textwidth}{>{\RaggedRight\arraybackslash}X >{\Centering\arraybackslash}X >{\Centering\arraybackslash}X >{\Centering\arraybackslash}X >{\Centering\arraybackslash}X >{\Centering\arraybackslash}X >{\Centering\arraybackslash}X}
\hline
\textbf{Work} &
\textbf{Platform} &
\textbf{Scalable} &
\textbf{Sparse} &
\textbf{Memory-Aware} &
\textbf{Interface-Aware} &
\textbf{Sim} \\

\hline
Eyeriss \cite{eyeriss2017} &
ASIC &
Limited &
No &
Partial &
No &
ASIC Simulation \\

Eyeriss v2 \cite{eyerissv2} &
ASIC &
Yes &
Yes &
Yes &
Limited &
ASIC Simulation \\

EIE \cite{eie2016} &
ASIC &
No &
Yes &
Yes &
No &
ASIC Simulation \\

Cnvlutin \cite{cnvlutin2016} &
ASIC &
No &
Yes (Activations) &
No &
No &
Simulation \\

FINN-R \cite{finnr2018} &
FPGA &
Limited &
Quantization Only &
Partial &
Yes &
FPGA \\

Pflow \cite{pflow2024} &
FPGA &
Limited &
No &
Yes &
Yes &
FPGA \\

VCONV \cite{vconv2024} &
FPGA &
Limited &
No &
Yes &
Yes &
FPGA \\

Ultra-Low Power FPGA CNN \cite{ultralowpowerfpga2024} &
FPGA &
No &
Partial &
Yes &
Limited &
FPGA \\

Sparse FPGA CNN \cite{sparsefpga2025} &
FPGA &
Limited &
Yes &
Yes &
Yes &
FPGA \\

Fast Readout FPGA CNN \cite{fastreadoutfpga2023} &
FPGA &
No &
Partial &
Yes &
Yes &
FPGA \\

\textbf{OpenEye (this work)} &
FPGA / ASIC &
\textbf{Yes} &
\textbf{Yes} &
\textbf{Yes} &
\textbf{Yes} &
\textbf{FPGA} \\

\hline
\end{tabularx}
\centering
\caption{Comparison of OpenEye with related CNN accelerator architectures}
\label{tab:related_work}
\vspace{-1em}
\end{table*}

\subsection{Contributions}
This work presents OpenEye, a scalable FPGA-based accelerator with independently configurable processing elements and clusters that enable adaptation to diverse performance requirements and resource constraints. The architecture features a hierarchical cluster-based design with dedicated configurable routers for flexible dataflow patterns and native support for sparse data processing in both activations and weights. We demonstrate near-linear scaling of computing performance vs. FPGA resources across multiple configurations on a Xilinx ZU19EG FPGA clocked at 200 MHz and provide a comprehensive evaluation that explicitly accounts for data transmission overhead and memory access costs, revealing the impact of communication bottlenecks on effective throughput.

\section{Architecture of OpenEye}

The OpenEye accelerator introduced in this paper is designed as a scalable and sparsity-aware hardware architecture for efficient execution of deep NNs on FPGA platforms. The architecture is composed of multiple interconnected clusters of processing elements (PEs), each capable of performing convolutional, fully connected, and pooling operations. OpenEye emphasizes modularity, scalability, and efficient data movement to minimize memory access overheads. The top-level architecture (Fig.~\ref{topmodul}) consists of a serial front-end responsible for interfacing with external memory and streaming data transfer, and a parallel back-end composed of multiple OpenEye clusters that perform the computations. Each cluster contains a configurable number of PEs organized in a two-dimensional array, along with dedicated routers for flexible data routing of activations, weights, and partial sums. The architecture supports sparse data representations, allowing for efficient processing of zero-valued activations and weights without unnecessary computations. The design is highly parameterizable, enabling adaptation to different performance requirements and resource constraints by varying the number of clusters and PEs per cluster. In the following subsections, we provide a detailed description of the top-level module, cluster architecture, PE cluster organization, and individual PE design.

\subsection{Topmodule}

\begin{figure}[tbp]
\includegraphics[width=\textwidth]{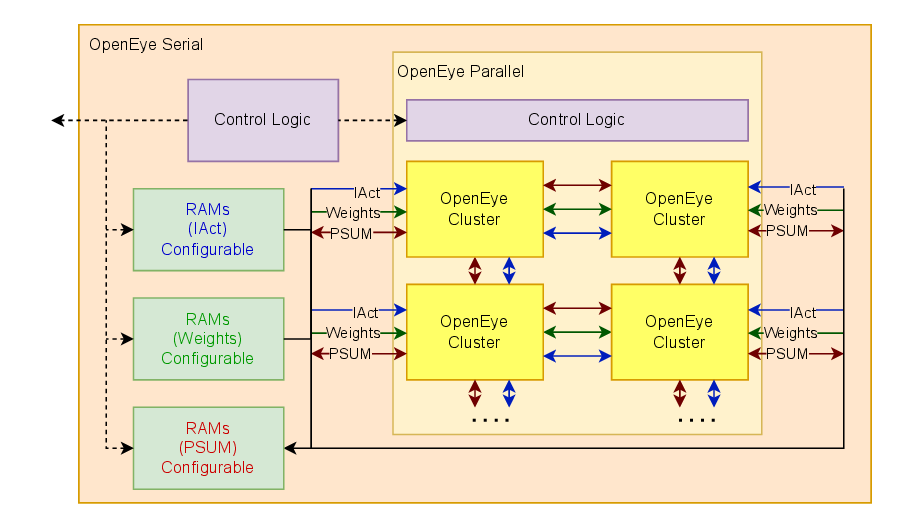}
\caption{Top-level architecture of OpenEye showing the configurable serial interface, centralized control logic, on-chip RAMs for activations, weights, and partial sums, and the scalable parallel cluster array.}
\label{topmodul}
\end{figure}

Fig.~\ref{topmodul} illustrates the top-level architecture of the OpenEye accelerator, which is divided into a serial front-end and a parallel compute back-end. The external interface of the serial module is fully configurable and is by default implemented using AXI or Wishbone with a 64-bit data port. This interface is responsible for streaming all network data and configuration parameters into the accelerator.

For the first network layer, the data stream contains the complete set of hyperparameters, input activations, weights, and biases. For all subsequent layers, only hyperparameters, weights, and biases are transmitted. This reduction in input bandwidth is enabled by the ability of OpenEye to store the complete intermediate feature maps of each layer in on-chip RAMs. Separate configurable RAM blocks are used for input activations, weights, and partial sums (PSUMs).

A centralized control logic orchestrates the execution of the network by iterating over all layers using a finite state machine. This control logic manages data movement, configures the routers within the individual OpenEye clusters, and synchronizes computation across the parallel accelerator array. The parallel module consists of multiple interconnected OpenEye clusters operating concurrently to achieve high throughput.

After computation is completed, the final results can be read back from the PSUM RAMs via the external interface. While the OpenEye parallel module can also be operated as a standalone design, e.g., for an ASIC-oriented approach, this would require more than 2000 external ports, making the serial front-end essential for practical system integration.

\subsection{OpenEye Cluster}
\begin{figure}
\includegraphics[width=\textwidth]{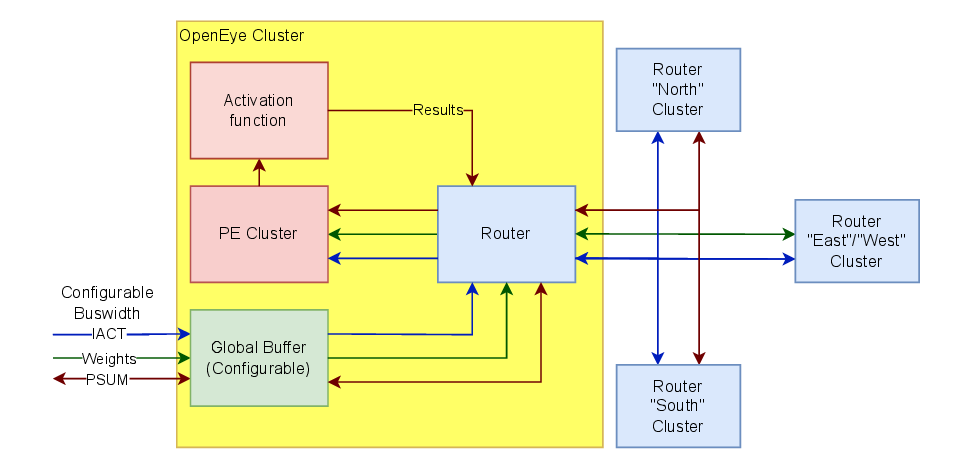}
\caption{Block diagram of an OpenEye cluster showing the PE cluster, configurable global buffer, activation function unit, and central router enabling scalable inter-cluster communication and flexible dataflow for sparse CNN acceleration.} \label{subcluster}
\end{figure}
Fig.~\ref{subcluster} shows the internal architecture of an OpenEye cluster, which represents the fundamental scalable building block of the accelerator. Each cluster comprises a PE cluster, an activation function unit, configurable global buffers, and multiple dedicated routers to enable flexible and scalable data movement. The global buffers are designed to be variable in both size and number, allowing adaptation to different network and layer requirements; however, the current implementation does not instantiate global buffers.

Each cluster integrates separate routers for input activations, weights, and partial sums. These routers are interconnected with corresponding routers of neighboring clusters, enabling structured inter-cluster communication. Input activations can be freely routed across clusters, while weight data is restricted to horizontal communication and partial sums to vertical communication. This directional dataflow reflects the computational structure of convolutional layers and supports efficient accumulation patterns.

Input activations and weights are transmitted unidirectionally, whereas partial sums are exchanged bidirectionally to enable accumulation across multiple clusters. Router configurations are centrally controlled by the top-level module, allowing the dataflow to be adapted dynamically to different network mappings. Data transfers between source and destination follow a ready/enable handshake protocol, ensuring robust synchronization and efficient flow control. Overall, the cluster architecture emphasizes modularity, scalability, and structured data movement to support sparse and efficient neural network acceleration.

\subsection{Processing Element Cluster}
\begin{figure}
\includegraphics[width=\textwidth]{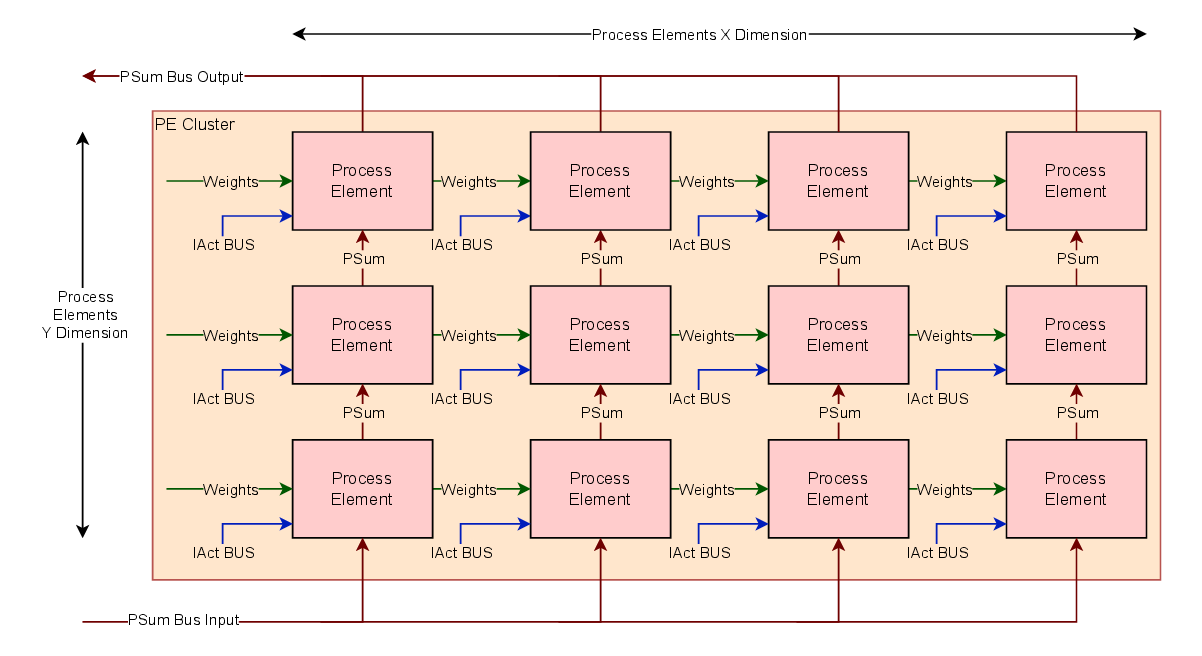}
\caption{PE cluster organized as a scalable two-dimensional array with independent X- and Y-dimension scaling and structured dataflow for activations, weights, and partial sums.} \label{PE_Cluster}
\end{figure}
Fig.~\ref{PE_Cluster} depicts the architecture of the OpenEye PE cluster, which forms the computational core of the accelerator. The PE cluster is organized as a two-dimensional array of PEs and is fully scalable along both the X and Y dimensions. This allows the degree of parallelism to be adapted to different performance requirements, resource constraints, and neural network layers.

Within the PE cluster, input activations are distributed via a dedicated activation (IACT) bus, while weights are propagated horizontally between neighboring PEs. Partial sums (PSUMs) are accumulated along the vertical direction using dedicated PSUM buses. This structured dataflow enables efficient convolutional processing and supports spatial reuse of both activations and weights.
Multiple PE clusters can be interconnected through the routers of the surrounding OpenEye clusters, allowing the composition of larger logical compute arrays. This enables the realization of larger convolution kernel sizes by spatially distributing the computation across multiple PE clusters. Furthermore, the routing of input activations is fully configurable, enabling diagonal data transmissions in addition to standard horizontal and vertical paths. Through this flexible configuration, input activations can also be streamed with arbitrary strides, allowing efficient mapping of different convolution parameters without modifying the underlying hardware. These features make the PE cluster a key contributor to the scalability and flexibility of the OpenEye architecture.

\subsection{Processing Element}
\begin{figure}[tbp]
\includegraphics[width=\textwidth]{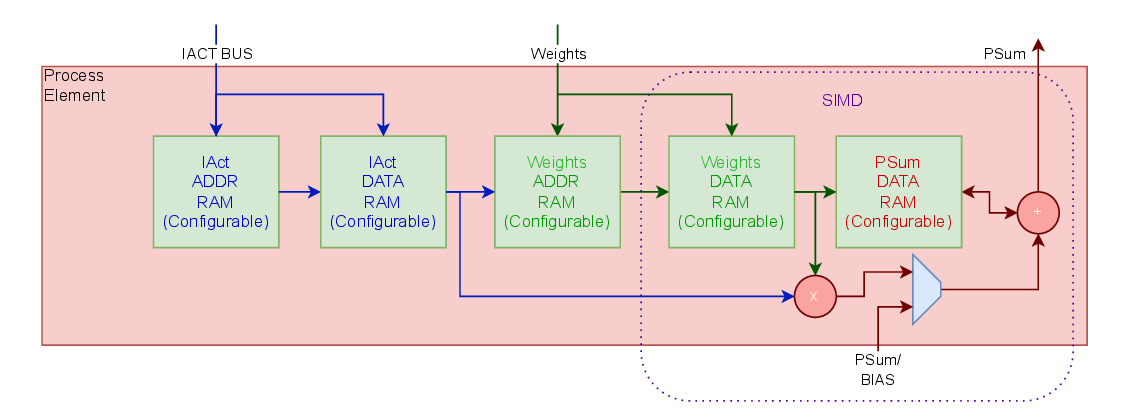}
\caption{Internal architecture of a PE showing sparse activation and weight decoding, hierarchical partial-sum accumulation with bias initialization, parameterizable SIMD execution, and FPGA-efficient PSUM storage using a Live Value Table (LVT) approach.} \label{PE}
\end{figure}

Fig.~\ref{PE} illustrates the internal structure of a single OpenEye PE. Input activations and weights are transmitted in sparse form and are subsequently encoded into dedicated address and data RAMs. Separate configurable RAMs are used for activation addresses and data, as well as for weight addresses and data, enabling efficient decoding of sparse input streams and reducing unnecessary memory accesses and computations. Detailed information about the sparse encoding scheme and its implementation can be found in the Eyeriss v2 paper~\cite{eyerissv2}, which serves as the basis for our sparse data handling approach and is omitted here for brevity.

The PE performs multiply–accumulate operations using locally stored activations and weights to compute Partial sums (PSUMs). PSUMs can be received via the lower input ports from vertically adjacent PEs, allowing hierarchical accumulation across the PE cluster. For the lowest PE in each column, the initial PSUM is initialized with the corresponding bias value, which is provided by the global buffers.
The degree of parallelism within a PE can be increased through parameterization to support SIMD-style execution. This parameterization scales the number of multipliers and adders and increases the width of the weight data RAMs accordingly, enabling higher throughput without modifying the overall architecture.

For the FPGA implementation, the PSUM data RAMs employ a Live Value Table (LVT)~\cite{laforest2012multi} approach, allowing logically multi-ported memory access to be realized using dual-port RAMs. This technique enables efficient accumulation of partial sums while remaining compatible with FPGA memory constraints.

\section{Evaluation}

To evaluate the scalability and performance of the OpenEye accelerator, multiple configurations were implemented on a Xilinx ZU19EG FPGA. The configurations varied in terms of the number of clusters and PEs per cluster to assess resource utilization, execution latency, and scalability behavior across different design points.

Beyond raw compute throughput, the evaluation explicitly targets the impact of architectural scaling on the end-to-end execution of convolutional neural network layers. In contrast to idealized accelerator evaluations that assume negligible communication overhead, OpenEye is designed to process complete layers within a single transmission cycle while minimizing accesses to external memory resources. Consequently, both computation and data movement are inherently reflected in the measured execution times.

The evaluated configurations scale the architecture along both spatial dimensions by varying the number of PE clusters and the internal PE array sizes, allowing an analysis of how parallelism, routing complexity, and memory structures interact at different scales. For each configuration, a representative convolutional neural network is executed on the hardware, and the total layer execution time is measured, including data ingestion, computation, and result storage.

All configurations were clocked at 200 MHz, and representative neural network operations were executed to analyze performance. The operations included are shown in Table~\ref{tab:cnn_architecture} and consist of a series of convolutional, pooling, and fully connected layers typical of a CNN architecture used for MNIST classification. The performance metrics collected for each configuration include throughput (measured in MOPS/GOPS) and resource utilization (measured in LUTs, FFs, BRAMs, and DSPs). These metrics were used to evaluate the trade-offs between performance and resource consumption across the different configurations, providing insights into the scalability behavior of the OpenEye architecture under varying design parameters.

\begin{table}[tbp]
\centering
\begin{tabular}{@{}llll@{}}
\toprule
\textbf{Layer Type} & \textbf{Filters/Units} & \textbf{Output Shape} & \textbf{Description} \\ \midrule
Input               & -                        & (28, 28, 1)           & Grayscale Image \\
Conv2D (1)          & 16                       & (28, 28, 16)          & 3x3 kernel, same padding \\
MaxPooling2D (1)    & -                        & (14, 14, 16)          & 2x2 pool, stride 2 \\
Conv2D (2)          & 32                       & (14, 14, 32)          & 3x3 kernel, same padding \\
MaxPooling2D (2)    & -                        & (7, 7, 32)            & 2x2 pool, stride 2 \\
Conv2D (3)          & 32                       & (7, 7, 32)            & 3x3 kernel, same padding \\
Flatten             & -                        & (1568)                & $7 \times 7 \times 32$ neurons \\
Dense (1)           & 32                       & (32)                  & Fully connected layer \\
Dense (2)           & 10                       & (10)                  & Output layer (Classification) \\ \bottomrule
\end{tabular}
\caption{Architecture of the Convolutional Neural Network}
\label{tab:cnn_architecture}
\end{table}
For the performance evaluation, an 8-bit quantized convolutional neural network targeting the MNIST dataset is employed. The network consists of three convolutional layers with interleaved max-pooling operations, followed by two fully connected layers for classification, as summarized in Table~\ref{tab:cnn_architecture}. All convolutional layers use $3 \times 3$ kernels with same padding, enabling a consistent spatial mapping onto the OpenEye PE clusters.

\begin{figure}
\includegraphics[width=\textwidth]{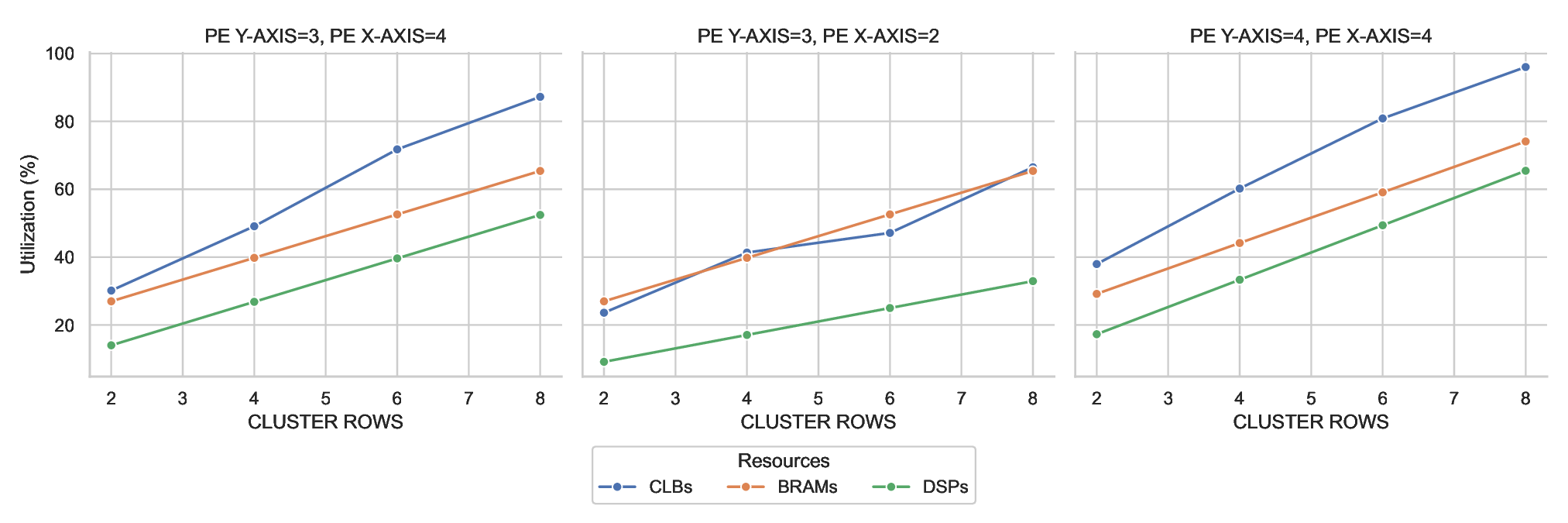}
\caption{FPGA resource utilization (CLBs, BRAMs, and DSPs) as a function of the CLUSTER ROWS parameter. Results are shown for three distinct configurations of PE X-Axis (PSUM) and PE Y-Axis (Weight). The Clock frequency is set to 200 MHz for all configurations.}
\label{Ressource_Usage}
\end{figure}

The network processes grayscale input images of size $28 \times 28$ and gradually increases the channel dimension to evaluate both spatial and channel-wise parallelism. With an overall computational complexity of approximately 2.13 million operations per inference, the model provides a representative workload that is sufficiently complex to stress the dataflow, routing, and memory hierarchy of the accelerator, while remaining compact enough to allow exhaustive evaluation across multiple hardware configurations.
This network is therefore well suited to analyze the execution latency, scalability behavior, and resource efficiency of OpenEye under realistic inference conditions.
Figure~\ref{Ressource_Usage} illustrates the FPGA resource utilization as a function of the number of cluster rows for three different PE array configurations. Across all evaluated design points, the utilization of CLBs, BRAMs, and DSPs increases in a strictly linear manner with the number of instantiated clusters. This behavior closely matches the architectural expectations and confirms that the OpenEye design scales predictably without introducing superlinear overheads due to routing, control logic, or memory structures.

The leftmost configuration corresponds to a PE organization comparable to that employed in Eyeriss v2, enabling a direct qualitative comparison with established spatial accelerator designs. Similar to prior work, increasing spatial parallelism primarily affects DSP utilization, which emerges as the dominant limiting resource for larger cluster deployments. In contrast, CLB and BRAM utilization exhibit more moderate growth, indicating that control logic, routing infrastructure, and buffering structures introduce only limited additional overhead when scaling the architecture.

Importantly, no visible inflection points, plateaus, or discontinuities can be observed in any of the resource trends. This indicates that neither routing congestion nor BRAM fragmentation dominates the scaling behavior within the evaluated design space. The absence of such effects suggests that the hierarchical routing structure and memory organization remain well-balanced even at larger cluster counts.

Notably, the consistent linear trends across all three PE configurations demonstrate that variations in the PE X- and Y-dimensions do not negatively impact scalability. This confirms that the underlying router-based interconnect and cluster composition strategy remain efficient even for asymmetric PE array shapes. Overall, these results validate the suitability of OpenEye for systematic design space exploration, as hardware resources can be accurately predicted and scaled according to performance requirements.

\begin{table}[tbp]
\centering
\small
\begin{tabular}{ C{1.7cm}  C{1.4cm}  C{1.4cm}  C{1.5cm}  C{1.5cm}  C{1.5cm}  C{1.5cm}  C{1.5cm}}
\toprule
\textbf{CLUSTER ROWS} & \textbf{PEs X-Axis} & \textbf{PEs Y-Axis} & \textbf{Data Send (ns)} & \textbf{Proc. Time (ns)} & \textbf{Total Time (ns)} & \textbf{MOPS (Proc.)} & \textbf{MOPS (Total)} \\
\midrule
1 & 2 & 3 & 70680 & 228635 & 299315 & 9330 & 7127 \\
2 & 2 & 3 & 106720 & 124545 & 231265 & 17127 & 9224 \\
4 & 2 & 3 & 131235 & 71475 & 202710 & 29844 & 10523 \\
8 & 2 & 3 & 132995 & 44525 & 177520 & 47908 & 12016 \\
\midrule
1 & 4 & 3 & 71960 & 127270 & 199230 & 16761 & 10707 \\
2 & 4 & 3 & 83680 & 70325 & 154005 & 30332 & 13851 \\
4 & 4 & 3 & 85225 & 42785 & 128010 & 49857 & 16664 \\
8 & 4 & 3 & 85580 & 29760 & 115340 & 71677 & 18494 \\
\midrule
1 & 2 & 4 & 82785 & 223310 & 306095 & 9552 & 6969 \\
2 & 2 & 4 & 130660 & 122020 & 252680 & 17482 & 8442 \\
4 & 2 & 4 & 162355 & 70180 & 232535 & 30395 & 9173 \\
8 & 2 & 4 & 163135 & 48745 & 211880 & 43761 & 10068 \\
\midrule
1 & 4 & 4 & 84045 & 121060 & 205105 & 17620 & 10400 \\
2 & 4 & 4 & 99920 & 67540 & 167460 & 31583 & 12738 \\
4 & 4 & 4 & 100985 & 41380 & 142365 & 51550 & 14983 \\
8 & 4 & 4 & 99915 & 29250 & 129165 & 72927 & 16515 \\
\bottomrule
\end{tabular}
\caption{Performance results for different OpenEye configurations.}
\label{tab:openeye_results}
\end{table}

\begin{figure}[!t]
\centering
\includegraphics[width=0.8\textwidth]{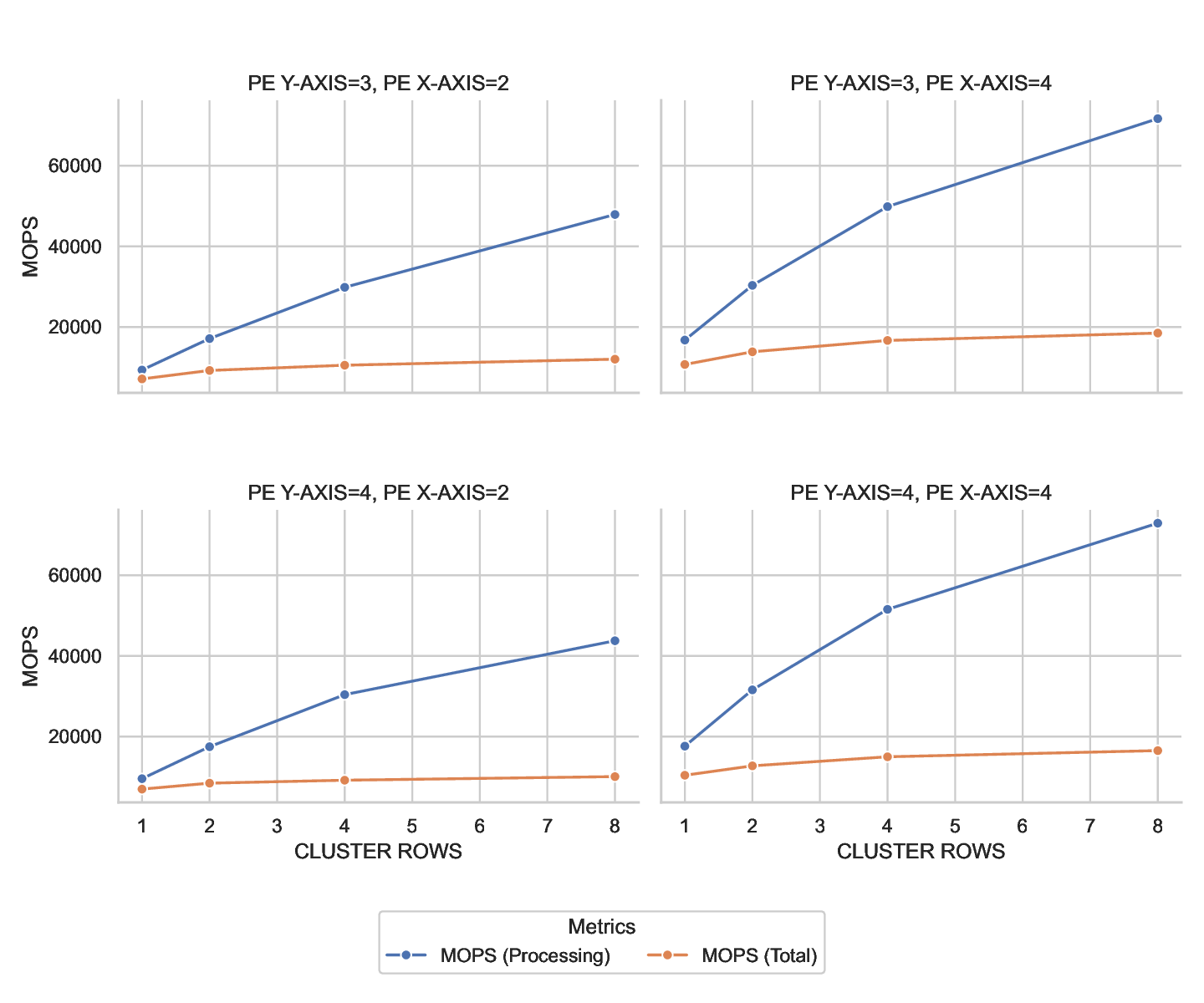}
\caption{Inference time for different configurations of the OpenEye}
\label{Time}
\end{figure}

Table~\ref{tab:openeye_results} and Figure~\ref{Time} summarize the performance results of the evaluated OpenEye configurations using the 8-bit quantized MNIST network. As expected, increasing the available hardware resources leads to a consistent reduction in execution time and a corresponding increase in computational throughput. This trend holds across all evaluated PE array shapes and cluster counts.

A closer inspection of the results reveals that the achieved processing throughput scales approximately proportional to the number of instantiated clusters. This behavior confirms that the spatial parallelism exposed by OpenEye is effectively exploited by the network and that the workload can be distributed efficiently across multiple clusters. However, the scaling efficiency gradually decreases for larger configurations. While the raw processing throughput (MOPS Processing) continues to improve near-linearly, the total system throughput (MOPS Total) exhibits diminishing returns.

This divergence is primarily caused by the increasing impact of data transmission overhead. As the number of clusters grows, a larger fraction of the total execution time is spent on data movement, reducing the effective duty cycle of the PEs. Consequently, fewer PEs are fully utilized at any given time, and the communication infrastructure becomes the dominant bottleneck for overall performance.

Furthermore, increasing the number of PEs along the Y-axis provides only limited performance benefits for the evaluated network. Since the workload is dominated by $3 \times 3$ convolutional layers, additional vertical PE parallelism cannot be fully exploited. Only the fully connected layers benefit from an increased number of PEs along this dimension, resulting in a comparatively small overall impact on throughput. This observation highlights the strong interaction between network topology and architectural scaling decisions and underlines the importance of workload-aware design space exploration.

\section{Conclusions and Outlook}

This work presented OpenEye, a scalable and sparsity-aware hardware accelerator for convolutional neural networks, designed to explicitly account for data transmission overhead and memory access costs—two aspects that are often idealized or neglected in accelerator evaluations. By enabling the execution of complete network layers within a single transmission cycle and minimizing external memory accesses, OpenEye addresses one of the dominant contributors to latency and energy consumption in modern deep learning workloads.

The architecture combines a modular cluster-based design with configurable routing, sparse data streams, and scalable processing element arrays. Experimental results on a Xilinx ZU19EG FPGA demonstrate strictly linear scaling of hardware resources with increasing cluster counts, confirming the architectural scalability predicted during design. No observable resource plateaus or inflection points were detected, indicating that routing congestion and BRAM fragmentation do not dominate within the evaluated design space.

Performance evaluation using an 8-bit quantized MNIST CNN shows that raw processing throughput scales near-ideally with additional clusters. However, the results also reveal that data transmission increasingly dominates total execution time, limiting effective throughput at larger scales. This effect is further reflected in the diminishing returns of increasing the PE Y-dimension for predominantly 3×3 convolutional workloads, where only dense layers benefit significantly from additional vertical parallelism.

In the future, we will evaluate the OpenEye architecture on a broader range of neural network topologies, including larger and more complex models with varying convolutional kernel sizes and channel dimensions. This will allow us to further analyze the interaction between network structure, architectural scaling, and communication overhead. Additionally, we will explore the impact of different sparsity patterns and quantization schemes on performance and energy efficiency.
Furthermore, the current findings motivate a systematic Design Space Exploration (DSE) as a natural next step. Future work will focus on jointly optimizing cluster dimensions, PE array aspect ratios, sparsity exploitation, and communication scheduling under both performance and energy constraints. In particular, OpenEye's modular structure makes it well-suited for automated DSE targeting application-specific workloads and deployment platforms, ranging from FPGA-based prototypes to ASIC implementations.

%
% Acknowledgements
%
\section*{Acknowledgements}
This work was funded by the German Federal Ministry of Research, Technology and Space (BMFTR) as part of the project FEntwumS -- Development Environment for the Digital World of Tomorrow, which is part of the federal funding program ``Digital Technologies'' -- under research grant number 16ME0984. We would like to thank our project partners for their support and collaboration, as well as the open-source community for their contributions to the tools and resources that made this research possible.

\paragraph{Competing interests.}
The authors declare that they have no competing interests.

%
% Bibliography
%
\bibliographystyle{unsrt}
\bibliography{reference}

\end{document}